# Navigating the Path of Women in Software Engineering: From Academia to Industry

Tatalina Oliveira
CESAR School
Recife, PE, Brazil
tcso@cesar.org.br

Ann Barcomb
University of Calgary
Calgary, AB, Canada
ann@barcomb.org

Ronnie de Souza Santos
University of Calgary
Calgary, AB, Canada
ronnie.desouzasantos@ucalgary.ca

Helda Barros
CESAR School
Recife, PE, Brazil
hbo@cesar.org.br

Maria Teresa Baldassarre
Università di Bari
Bari, BA, Italy
mariateresa.baldassarre@uniba.it

César França
CESAR School
Recife, PE, Brazil
franssa@cesar.org.br

## ABSTRACT

*Context.* Women remain significantly underrepresented in software engineering, leading to a lasting gender gap in the software industry. This disparity starts in education and extends into the industry, causing challenges such as hostile work environments and unequal opportunities. Addressing these issues is crucial for fostering an inclusive and diverse software engineering workforce. *Aim.* This study aims to enhance the literature on women in software engineering, exploring their journey from academia to industry and discussing perspectives, challenges, and support. We focus on Brazilian women to extend existing research, which has largely focused on North American and European contexts. *Method.* In this study, we conducted a cross-sectional survey, collecting both quantitative and qualitative data, focusing on women's experiences in software engineering to explore their journey from university to the software industry. *Findings.* Our findings highlight persistent challenges faced by women in software engineering, including gender bias, harassment, work-life imbalance, undervaluation, low sense of belonging, and impostor syndrome. These difficulties commonly emerge from university experiences and continue to affect women throughout their entire careers. *Conclusion.* In summary, our study identifies systemic challenges in women's software engineering journey, emphasizing the need for organizational commitment to address these issues. We provide actionable recommendations for practitioners.

**LAY ABSTRACT**. Women in software engineering confront enduring underrepresentation, creating a gender gap evident from education to industry. Increasing the participation of women in the software industry is essential for the development of systems that reflect the diversity of our society. This study explores women's journey within software engineering, highlighting challenges and proposing strategies for software companies to address gender imbalance effectively.

## KEYWORDS
women, software development, STEM, survey, gender

## 1 INTRODUCTION

Throughout the years, there has been a rise in research efforts aimed at tackling the issue of equity, diversity, and inclusion in the field of software engineering [51]. Studies conducted in both educational and industry settings have consistently revealed that software engineering remains unwelcoming to individuals from various underrepresented groups [2, 16]. Disturbingly, these individuals face various forms of prejudice, ranging from toxic work environments to deliberate discrimination [15, 17, 55]. The effects of this imbalance extend to the software products being released to our naturally diverse society.

In this context, women continue to be significantly underrepresented in software engineering [67]. Despite progress in promoting gender diversity in various industries, the software industry continues to grapple with a substantial gender gap. This underrepresentation starts from early education, with a small percentage of girls pursuing computer science and engineering degrees compared to their male counterparts [47]. This lack of gender diversity deprives the industry of diverse perspectives and perpetuates a discouraging cycle that dissuades potential female talent from pursuing or maintaining careers in software engineering [26, 27].

In the academic environment, beyond the issue of underrepresentation, girls pursuing software engineering may encounter many challenges: exposure to gender stereotypes; experiencing a lack of mentorship tailored to their needs; feeling as if they do not belong; having limited access to internships, research projects, or leadership positions; suffer from the absence of diverse and inclusive curriculum; and feel discomfort in male-dominated classrooms [34, 44, 45, 47].

After entering the software industry, certain challenges may persist, including underrepresentation and male-dominated teams. However, women encounter a range of additional obstacles, such as doubting their abilities and contributions (impostor syndrome), contending with hostile work environments, and facing a lack of recognition and unequal pay. Additionally, balancing work demands with family responsibilities can be particularly challenging for women, affecting their career choices and advancement afterward [26, 35, 66–68].

Building upon the research on this subject, we aim to extend and enrich the existing literature regarding women's experiences in the field of software engineering and explore their trajectory from the university to the industry, discussing the obstacles encountered, the support received, and how they perceive the potential for motivating other women to enter the field. Because the majority of studies either do not specify the population or focus specifically on Western nations, we opted to explore the paths of Brazilian women navigating the software industry. Brazil's intricate sociocultural landscape, characterized by a heterogeneous population encompassing varied ethnicities, cultures, and social norms, offers an excellent environment for studying gender diversity and inclusion within software engineering. To address this topic, we focused on the following research question: *What is the narrative of Brazilian women's experiences in software engineering, and how do they describe their journey into the field?*

From this introduction, our study is organized as follows. Section 2 examines existing research on women in software engineering. In Section 3, we outline our methodology, while Section 4 presents the findings. Section 5 delves into the implications and limitations of our study. Lastly, Section 6 summarizes the contributions made by this research.

## 2 BACKGROUND

Noteworthy achievements, obstacles, and advancements characterize women's history in software engineering [5]. During the early days of computing, women played vital roles as programmers, making significant contributions to the field [7]. Despite their contributions, women have often not been given credit for their work, or their efforts have been less accepted [9, 63, 64]. Over the decades, the representation of women in software engineering gradually diminished primarily due to cultural and societal factors, perpetuating the misconception that programming is better suited for men and discouraging girls from entering the field [2, 12, 60, 64]. Stereotypes of the programmer as someone with a consuming interest in programming and poor social skills also discourage people who do not see themselves represented in this picture from entering the field or from applying for particular jobs [12, 21, 52, 69].

As students, women are widely exposed to an environment where "sucking it up" is normalized [4, 71]. At the same time, stresses related to being marginalized can exacerbate mental health conditions [14]. Such negative experiences during formative years can lead to impostor syndrome and negative self-talk, potentially contributing to sabotaging career behaviors such as avoiding opportunities due to feelings of inadequacy [40, 41].

Several stereotypes persist about women, which can lead to discrimination and bias. One common stereotype is that women are less competent with technology, while another is that women prioritize family life over their work [1, 33, 42, 58, 70, 72]. These stereotypes can also be internalized, leading to women making career decisions based on self-assessed competency [33, 57, 58]. In the software industry, women may face harassment in various forms, such as sexual harassment and being ignored, which can lead to feelings of exclusion [36, 38, 65]. Discrimination regarding pay and promotions can also affect women's careers, and their retention in the field [13, 24, 28, 38, 56, 64]. If women choose to become mothers, they are further judged as inadequate workers and bad mothers, and may see responsibilities and opportunities reduced [24, 25, 43]. As women grow older, they may further find themselves excluded by a perceived lack of sexual attractiveness [19, 68].

Yet, women make substantial contributions to software development, using their distinctive personalities, strong problem-solving abilities, and innate creativity to bring valuable perspectives and innovation to the industry [10, 53]. Their technical expertise makes them valuable assets in roles like programming, software architecture, quality assurance, and project management. In addition, women's advocacy for diversity and inclusion leads to innovative solutions that cater to a wide range of user needs [32, 53, 59, 73].

In this sense, shedding light on the challenges confronted by women in the software industry and putting forth compelling strategies to foster gender parity in software development is crucial these days to ensure equal opportunities for women in the tech industry and support the continuous growth and long-term success of the software development field [10, 26, 67]. Our work aims to add to the rich body of literature on women's experiences in the software industry by looking at the transition from education to employment and by focusing on the experience of women beyond the western countries that inform the majority of existing research.

## 3 METHOD

In this research, we conducted a cross-sectional survey [20] to explore how women navigate software engineering, from academia to their first job to their experiences in everyday software development projects. To design our research strategy, we followed three guidelines for conducting surveys in software engineering [37, 48, 49]. Below, we provide a detailed account of each step in our methodology. The following sections detail our method, which is illustrated in Figure 1.

### 3.1 Building up on the Literature

Many studies have been recently published addressing the issue of gender imbalance in software engineering, often focusing on either the academic context or the industrial setting. Our research aimed to build upon the existing literature by exploring women's journey in software engineering from university to software companies. By investigating their perceptions, challenges, and motivations in both settings, we followed the narratives of a sample of women to enrich the understanding of this problem. Through this approach, we seek to extend the knowledge available and gain valuable insights to address the gender gap in the field.

### 3.2 Instrument

Our survey questionnaire was crafted by drawing upon numerous published studies concerning women's experiences in software engineering in academia and the industry. We designed a comprehensive mix of qualitative and quantitative questions to capture our participants' backgrounds and narratives.

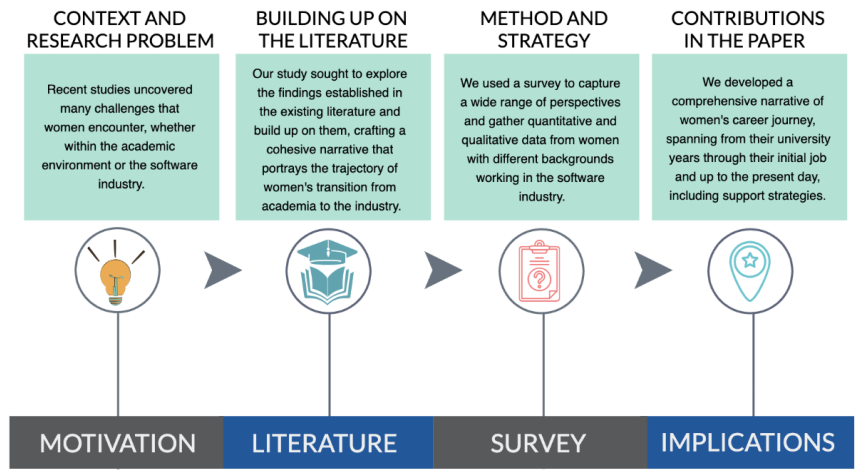

Figure 1: Method

To address the challenges women face in software engineering, which have been extensively discussed in existing literature, we initially requested participants to name and discuss obstacles encountered in their journey. Subsequently, we provided them with a list of challenges from prior studies, allowing them to select relevant ones based on their experience.

In addition, our questionnaire encompassed inquiries about their motivation to pursue software engineering, the level of support they received from academia and the industry, first job experiences, and current experiences in the software industry. Furthermore, we collected information about their roles in software development (e.g., their main tasks within their teams) to gain insights into how challenges might affect specific groups of women differently.

We conducted a validation process for the questionnaire involving three women in the software industry who completed the survey and provided valuable feedback for improvement. Based on their input, we determined that the average time to complete the survey was approximately 10 minutes. We made necessary enhancements, including rephrasing certain sentences and allowing closed-ended questions to be answered optionally. This modification was in response to feedback indicating that some women might not have experienced the challenges mentioned in the survey, despite being derived from literature findings. It is important to note that the participants of the pilot study were not included in our final sample. The revised and final version of the questionnaire is available in Table 1.

### 3.3 Data Collection

Following [3], we applied two sampling strategies in our data collection, namely, convenience sampling and snowballing. Convenience sampling is a method that involves selecting participants based on their availability to take part in the research. In contrast, snowball sampling relies on initial participants helping to identify and recruit additional participants, leading to the expansion of the participant network in an iterative manner.

We employed convenience sampling to identify women in the software industry who might be interested in participating in our study. Using our extensive network of software professionals, we directly sent the questionnaire to them via email, chat, and social media. Additionally, we advertised our questionnaire to women working at a prominent software company in Brazil that develops software for clients from various countries.

Furthermore, following snowball sampling, we requested the women from the convenience sample to forward our questionnaire to colleagues and co-workers who belong to our target population and might be interested in participating.

### 3.4 Data Analysis

In this study, both quantitative and qualitative data were collected, leading to the application of two distinct data analysis techniques: descriptive statistics [22] for quantitative data and coding techniques [23] for qualitative data.

To start, we utilized descriptive statistics [22] to describe the main characteristics of our sample quantitatively. By partitioning participants' responses into sub-groups using various statistical functions (e.g., means, proportions, totals, ratios), we gained insights into the data. We employed Tableau to create interactive visualizations, facilitating the analysis of data patterns such as averages and distributions. Aggregations were used to explore the experiences of different groups of women in the sample based on factors like roles and backgrounds.

Subsequently, we applied qualitative analysis to open-ended questions in the questionnaire, involving three coding stages: line-by-line, focused, and theoretical coding [23]. With line-by-line coding, we carefully examined each participant's responses to construct codes that reflected their individual experiences and perspectives. Moving on to focused coding, we identified connections among the data and created high-level categories that categorized the women's trajectory in software engineering. Finally, through theoretical coding, we iteratively rearranged the categories to create a coherent

**Table 1: Survey Questionnaire**

[OPENING] Welcome to our research on women's journey in software engineering. This survey is COMPLETELY ANONYMOUS, and your responses cannot be linked to you. Please answer the questions based on your experience. The survey takes up to 10 minutes to complete. Will you participate?
( ) Yes

1. Reflecting on your college experience, could you share any challenges you encountered as a woman in software engineering/computer science that your male colleagues did not have to deal with?

2. While in college, were you aware of any women who made significant contributions to the field of computing? If so, please share your thoughts.

3. Did you have any female role models in the field who inspired you? If so, share your thoughts or comments on how they influenced your journey.

4. [OPTIONAL] During your time in college, did you encounter any of the following challenges? Please select all the options that apply:
( ) I felt insecure.
( ) I felt that was not my place.
( ) I heard jokes and stereotypes from classmates.
( ) I experienced moral harassment.
( ) I experienced sexual harassment.
( ) I lacked female role models to follow.
( ) I didn't have early exposure to computers and technologies like my male peers.
( ) I was afraid of not achieving career success.

5. What challenges did you face in your first job at a software development company that you felt were related to being a woman?

6. As a woman working in your first job in software engineering, what support or assistance did you receive from colleagues or managers that proved to be crucial for you?

7. As you think about your daily work on software projects today, tell us about or experience as a woman in the software industry.

8. As you perform your specific role or activity on the team (e.g., programmer, tester, designer, manager, etc.), could you share any obstacles you have encountered as a woman in this capacity? Please specify your role and describe the challenges faced.

9. [OPTIONAL] Considering the new remote and hybrid work structures, share your experience as a woman working in this context.

10. [OPTIONAL] Please consider the list below and select all the main challenges you have personally faced in your work in the software industry:
( ) I've felt invisible in projects because I'm a woman.
( ) I've felt excluded from conversations and decisions because I'm a woman.
( ) I've heard toxic comments or jokes from team members about being a woman.
( ) I felt that I was not contributing enough because I am a woman (imposter syndrome).
( ) I felt that I don't have support from the leaders because I am a woman.
( ) I've been treated as some kind of stereotype for being a woman.
( ) I felt like I could not balance my personal life, home, and work because I am a woman.
( ) I've felt my work was overlooked or undervalued because I am a woman.

11. What is your highest educational level?
( ) High-School
( ) Bachelor's degree
( ) Post-baccalaureate
( ) Master's degree
( ) Ph.D.

12. What is your primary role in the team?
( ) Software Requirements
( ) Software Design
( ) Software Management
( ) Software Programming
( ) Software QA
( ) Other role

13. Do you have small children at home?

14. What best describes your ethnicity?

15. Do you identify as a member of any other underrepresented group in software engineering?
( ) LGBTQIA+
( ) Neurodivergence
( ) PWD
( ) Senior (50+)
( ) None

16. Are you a trans woman?

[CLOSING] This survey guarantees complete anonymity, and your information will not identify you. However, we may conduct 30-minute interviews with some participants to enhance the study. If interested, please leave your e-mail address, and we'll contact you to schedule the meeting. Thank you for your valuable participation.

narrative based on the collective experiences of our sample. Interviews were conducted in Portuguese; quotations provided in this paper have been translated by one or more authors.

### 3.5 Ethics

To ensure the anonymity of participants, no personal information (e.g., name, e-mail, or employer) was collected in this study. As shown in Table 1, the beginning of the questionnaire included general information about the study and the research team, and participants were asked to give their consent for using their data for scientific purposes by checking a 'yes' box. In the end, only those interested in participating in our subsequent research voluntarily provided us with their email.

## 4 FINDINGS

In this section, we present the findings obtained from our preliminary analysis. We begin by summarizing the characteristics of our participant sample. Subsequently, we demonstrate how women experience software engineering across three stages of their careers: academia, first job experience, and daily work in the industry.

### 4.1 Demographics

We collected 42 valid questionnaires from a diverse group of women working in software engineering with various profiles and backgrounds. They are engaged in a wide range of activities in software development and were able to share various experiences in the field, offering both insights aligned with existing literature and novel perspectives. Table 2 provides a comprehensive sample overview.

In summary, our sample comprises highly skilled women, with 71% holding educational levels beyond an undergraduate diploma, including post-baccalaureate certificates, Master's, and Ph.D. degrees. Furthermore, our sample consists of women working across five areas of software development, with over 81% engaged in coding activities like programming and testing.

Apart from their educational and technical backgrounds, our sample includes women from other underrepresented groups in software engineering: 47% are non-white, 9% have neurodivergence, and 9% identify as part of the LGBTQIA+ community (note that no trans woman participated in the study).

Finally, a group of women (21% of the sample) are working mothers who must balance their professional responsibilities with caregiving duties.

### 4.2 Academic Life

In the academic setting, women face a series of obstacles that they must overcome to successfully graduate and enter the industry. Our survey identified several experiences in dealing with these challenges, as well as compelling ideas to tackle this issue. Below, Table 3 summarizes the perspective of women who have triumphed over challenges in software engineering courses and successfully transitioned to the industry.

#### 4.2.1 Challenges in Academia.

In technology courses, girls often battle the (FEAR OF FAILURE, primarily due to self-doubt and the predominantly male environment. This insecurity is amplified by enduring and persistent (JOKES and stereotypes directed at them during their university years. P18 reported, *"You need to deal with offensive and insulting comments that, unfortunately, are recurring."*

**Table 2: Survey Demographics**

| | | |
|---|---|---|
| Education | High-School | 2% (1 woman) |
| | Bachelor's degree | 26% (11 women) |
| | Post-baccalaureate | 45% (19 women) |
| | Master's degree | 19% (8 women) |
| | Ph.D. | 7% (3 women) |
| Role in SE | Software Requirements | 2% (1 woman) |
| | Software Design | 2% (1 woman) |
| | Software Management | 14% (6 women) |
| | Software Programming | 45% (19 women) |
| | Software QA | 36% (15 women) |
| Mother of Small Children | No | 79% (33 women) |
| | Yes | 21% (9 women) |
| Ethnicity | Mixed-race | 50% (21 women) |
| | White | 45% (19 women) |
| | Black | 2% (1 woman) |
| | Prefer not to answer | 2% (1 woman) |
| Diversity Group | Neurodivergence | 9.5% (4 women) |
| | LGBTQIA+ | 9.5% (4 women) |
| | Senior (50+) | 2% (1 woman) |
| | None | 2% (33 women) |

**Table 3: Academic Experiences**

| | | |
|---|---|---|
| Challenges | Insecurity | 62% (26 women) |
| | Scarcity of Female Role Models | 48% (20 women) |
| | Jokes and Stereotypes | 43% (18 women) |
| | Moral Harassment | 40% (17 women) |
| | Insufficient Exposure to Computers | 31% (13 women) |
| | Reduced Sense of Belonging | 31% (13 women) |
| | Distrust | 29% (12 women) |
| | Sexual Harassment | 10% (4 women) |
| | Opportunities Shortage | 10% (4 women) |
| | Career Assumption | 10% (4 women) |
| | Group Segregation | 5% (2 women) |
| Role Models | Teachers | 26% (11 women) |
| | Family Members | 14% (6 women) |
| | Industry Workers | 7% (3 women) |
| | Women in SE History | 5% (2 women) |
| | Other Minorities in SE History | 2% (1 woman) |
| | Company CEO (Men) | 2% (1 woman) |
| Solutions | Women Visibility | 21% (11 women) |
| | Female-affirmative Teaching Positions | 14% (6 women) |
| | Female-focused Activities | 12% (5 women) |
| | Support Networking and Channels | 12% (5 women) |
| | Career Discussions | 7% (3 women) |
| | Diversity Awareness | 5% (2 women) |
| | Early Engagement | 2% (1 woman) |

Another difficulty women encounter in software engineering courses is the scarcity of female role models. Despite the existence of women who have made noteworthy contributions to technology throughout the history of computer science, this narrative is often disregarded or overshadowed. P15 highlighted, *"I always missed a female figure (…) parameters and examples to follow, to know where I would end up."*

Additionally, the majority of professors in the field are men; consequently, girls find it challenging to envision themselves as part of the domain, leading to a reduced sense of belonging. Furthermore, this situation renders them susceptible to various forms of misconduct, including but not limited to moral harassment and sexual harassment. P30 reported having witnessed professors *"comparing students using beauty standards stereotypes."*

According to our participants, girls often have insufficient exposure to computers and technology development during their formative years when compared to boys, which subsequently influences and shapes their experiences within the course. P06 commented *"They would always tell stories about using computers for many things and already knowing more. I felt somewhat left out because everything I knew, I learned at the university.""*

Our findings demonstrate that group assignments pose a concern in courses dominated by male students, as girls often encounter difficulties in securing teammates, as they are unfairly linked to perceived weaknesses within the program (group segregation). P18 mentioned that *"we need to continuously keep proving ourselves to classmates and professors."*

Another significant challenge in this context is the persistent questioning of their ability to effectively tackle computational problems (distrust), which sometimes results in their unjust confinement to particular tasks within software development activities. P38 discusses that *"in group projects, I was always assigned with documentation because women are 'more organized' and guys are more 'logic-oriented' to deal with the code."*

This lack of confidence in young women's capabilities appears to impact not just their university years but also their future prospects, as our participants reported opportunities shortage, including internship selections and enduring gender-based career assumptions within the software industry. P07 commented, *"There was a "premise" that women are not competent to code. They always used to say, "You are in charge of the documentation.""*.

#### 4.2.2 Support in Academia.

Our participants understand that the primary approach to addressing these issues within the academic environment entails the creation of a network of academic support tailored for girls, e.g., a secure space where they can openly share and discuss the challenges they encounter in the program (support networking and channels). Additionally, it is essential to develop female-focused activities, facilitating interactions among peers in a welcoming setting (i.e., study groups, focused discussions, hackathons, and similar endeavors). P37 says that one solution could be based on *"creating affirmative opportunities for women, forming groups focused on women, and establishing a channel for reporting cases that exceed boundaries."*

Further, software engineering programs should strive to enhance diversity within their departments, ensuring stronger female-affirmative teaching positions, which can serve as sources of inspiration. P02 added that *"we need a greater women representation in the universities and promoting awareness and training within the academic community."* P17 reinforces: *"starting from the university itself, which should hire more women to teach, for instance."*

Another meaningful step towards addressing this concern in the academic environment is to raise diversity awareness in software development and enhance women['s] visibility by highlighting their significant contributions to the field. Engaging in career discussions about potential career paths and the various roles available within software companies can substantially aid girls in envisioning their involvement in this field. P24 discuss that we should *"introduce career possibilities to them that go beyond* *programming (e.g., tester, product owner, business analyst, project manager, UX, etc.).*"

Finally, there is a collective advantage in EARLY ENGAGEMENT through familiarizing both genders with successful instances of women in software engineering as well as comprehending the favorable outcomes that stem from diverse teams in software and technology outcomes. P14 complements by saying that *"showcasing more cases of women in the computing field and organizing events where female speakers take the stage"* are interesting strategies in this scenario.

### 4.3 Transitioning to Software Industry

As women transition from academia to their initial roles in the software industry, a set of issues faced at school persist, and some challenges faced before do not diminish entirely for 69% of our sample with the shift in the environment. However, the industrial context appears to be somewhat more receptive, with 31% of the respondents indicating that they encountered no major challenges in their initial job. In addition to challenges, our participants discussed various types of support received in their first job experience that was crucial for them to succeed in a software engineering career. Below, Table 4 summarizes the perspective of women who made it to industry and overcame the obstacles of their first job.

**Table 4: First Job Experiences**

| | | |
|---|---|---|
| Challenges | Distrust | 38% (16 women) |
| | Stress | 12% (5 women) |
| | Biased Hiring | 12% (5 women) |
| | Harassment | 7% (3 women) |
| | Work-life Balance | 5% (2 women) |
| Support | No Support | 33% (14 women) |
| | Team Empathy | 26% (11 women) |
| | Women in the Team | 19% (8 women) |
| | Technical Help | 14% (6 women) |
| | Regular Feedback | 7% (3 women) |

#### 4.3.1 Challenges in the First Job.

Looking at the challenges, women's competencies remain under scrutiny in their first industry experience, and some (male) co-workers consistently cast doubt on the quality of their work while diminishing their contributions to the software project (DISTRUST). P04 revealed: *"they didn't trust my work, and my doubts were always treated as something that didn't even deserve discussion."* P17 adds to the discussion by reporting that *"my knowledge was questioned, or my opinion invalidated, and this would often happen right after I expressed an opinion or idea, only for a male team member to appropriate it as if it were his own."*

Various forms of misconduct and toxic behaviors persist as women transition from academia to the industry, encompassing harassment, inappropriate jokes, and harmful stereotypes (HARASSMENT). Our respondents shared that women in male-dominated teams frequently find themselves navigating through uncomfortable situations, which not only heighten STRESS levels but also evoke familiar emotions experienced during academia, such as a low sense of belonging and increased feelings of insecurity. P02 remembers that she was *"in charge of a particular task, and the boy who was supposed to assist with its execution didn't want to talk to me, but only to my manager, to inquire about the issue and seek clarification."* P01 also experienced a highly toxic environment, as she narrates: *"I felt uncomfortable with jokes or overly intimate remarks coming from a technical leader, insinuating that I might be having an affair with the boss due to my promotion within the company."*

Our findings also demonstrated that transitioning from academia to the industry not only fails to alleviate the challenges faced by women but also introduces new obstacles that must be navigated. Our participants highlighted BIASED HIRING, with women frequently being assessed based on appearance or subjected to excessive testing with questions typically not posed to their male counterparts. P13 commented that *"in the job interview, my colleague later told me that I only got through because I wasn't wearing makeup. According to him, women who wear makeup can't code."*

Moreover, as they embark on their first jobs, women frequently find themselves contending with the immediate challenges of WORK-LIFE BALANCE and navigating critical career decisions, as highlighted by P05: *"they lead me to believe that having children and other household responsibilities would impact my performance as a professional in the field"*; and complemented by P39: *"I was 31 years old in my first job, married, and with a daughter, it was truly traumatic, and I felt the urge to give up everything."*

#### 4.3.2 Support in the First Job.

Our respondents highlighted four key factors that aided their transition into their initial job roles, empowering them to overcome challenges and establish themselves as accomplished software professionals. The majority of this support stems from collaborative and interpersonal facets, as well as effective managerial strategies. In contrast, only one element pertains to technical or technological support for embarking on their career journey.

The TEAM EMPATHY displayed by colleagues stands as a crucial component in aiding women to thrive in their first job opportunities in the software industry. Team members can foster an inclusive atmosphere by establishing internal norms that prioritize respect and firmly reject any form of discrimination and misconduct. P2 commented on empathy by stating: *"I was very fortunate to have extraordinary people by my side, who usually intervened and clarified my role within the project, repudiating the incidents."* P15 also experienced the power of empathy in her first job: *"My first manager always made sure that everyone respected each other (…) this is how I began to have more confidence in my potential."*

The presence of WOMEN IN THE TEAM plays a vital role in providing support for female software professionals. As previously mentioned, the absence of references and role models in academia poses a significant challenge for girls. Thankfully, the enhanced gender diversity within software companies can help bridge this gap in the early stages of their careers. P29's report illustrates this finding: *"I had two friends who graduated with me, and we always supported each other in making decisions; having that support definitely helped me a lot."* P30 added her thoughts on the subject: *"It was crucial when my manager was a woman, as she recognized my potential, and helped me to develop my skills."*

Some bad academic experiences, such as persistent capabilities questioning or being assigned to meaningless activities in group projects, can potentially instill a lasting sense of failure or low performance among girls. Consequently, our findings propose that REGULAR FEEDBACK can serve as a valuable aid in empowering them during the initial phases of their careers. P42 remembered her first job experience: *"performance feedback brought me more self-confidence."* This feeling was shared by P07: *"(...) positive feedback, and motivational conversations that reassured me of my capabilities."*

Finally, similar to any other newcomer to a software project or junior professional, girls also require technical guidance from their leaders to achieve success. As our results demonstrated in Section 4.3.1, there are instances where women are overlooked by their (male) colleagues, undermining their prospects of accomplishing their tasks. Thus, TECHNICAL HELP remains a crucial factor. P16 provided an example on this matter: *"I received help and support in what I needed to become a better professional and deliver better results. For instance, I engaged in pair programming to enhance my skills and advance my career."*

## 4.4 Everyday in the Software Industry

Our final set of findings encompasses the ongoing challenges that women continue to encounter in software development, alongside strategies that can be employed to overcome these obstacles. Both these challenges and solutions arise from the real-life experiences of female software professionals who have successfully navigated through a biased and, at times, hostile academic setting and a challenging first job experience.

In other words, they managed to succeed and become well-established software professionals. Yet, this achievement doesn't imply that challenges have vanished or belong to the past. Software engineering remains notably lacking in diversity to improve women's experience. Table 5 summarizes the findings regarding the everyday experiences of women in the software industry.

**Table 5: Current Experiences**

| | | |
|---|---|---|
| General Challenges | Impostor Syndrome | 28% (67 women) |
| | Invisibility | 52% (22 women) |
| | Jokes and Stereotypes | 50% (21 women) |
| | Exclusion from Decisions | 50% (21 women) |
| | Work-life Balance | 48% (20 women) |
| | Toxicity | 48% (20 women) |
| | Distrust | 31% (13 women) |
| | Undervalued Work | 29% (12 women) |
| | Lack of Support | 24% (10 women) |
| | Loneliness | 14% (6 women) |
| | Unequal Salary | 12% (5 women) |
| Solutions and Mitigation | Diversity Awareness | 26% (11 women) |
| | Female Leadership | 17% (7 women) |
| | Team Empathy | 12% (5 women) |
| | Professional Development | 10% (4 women) |
| | Team Diversity | 7% (3 women) |
| | Team Inclusiveness | 7% (3 women) |
| | Unbiased evaluation | 5% (2 women) |
| | Whistleblowing Structures | 5% (2 woman) |
| | Childcare Support | 2% (1 woman) |
| | Egalitarian Salary | 2% (1 woman) |
| | Misconduct Policies | 2% (1 woman) |

### 4.4.1 Everyday Challenges.

As demonstrated in Table 5, numerous challenges have persisted for women in software engineering from their university experiences, continuing to pose difficulties in their day-to-day work. Elements such as JOKES AND STEREOTYPES, HARASSMENT, INVISIBILITY, LACK OF SUPPORT, and DISTRUST about their competence are factors that, as our findings suggest, do not merely dissipate over time. In fact, certain challenges appear to manifest in more pronounced and detrimental ways.

The continuous disregard and skepticism towards women's abilities in software development have contributed to the development of IMPOSTOR SYNDROME among many, a condition that can eventually lead to various health concerns over time. P18 commented that: *"(...) emotional concerns like anxiety, impostor syndrome, and insecurity appear to be common among all women in the field."*

In addition to the aforementioned and previously described issues, women have frequently encountered a sense of isolation at the workplace (LONELINESS), attributed to the scarcity of female colleagues within their teams. It's important to note that increasing the female workforce in a company does not necessarily equate to achieving gender parity in software projects. In certain instances, female software professionals may find themselves working on projects for extended periods being the only woman on the team. P41 commented on this issue: *"I think there's a bit of a lack of a sense of belonging. (...) I still don't know how to initiate a friendship when I meet another woman in my field."* P02 also added to this topic: *"In my current job, I did not face major issues so far; however, I miss having more women coding in the teams."*

Another challenge encountered by female software professionals as they advance in their careers is the pervasive problem of UNEQUAL SALARY, a concern observed in numerous industries, including software engineering. Our participants raise awareness about this challenge and emphasize the obstacles women encounter when seeking promotions or salary increments. P13 stated that *"there is a challenge in being promoted, achieving equal salaries, and attaining leadership positions."*

Still, regarding this issue, female software professionals have encountered instances where their ideas were disregarded and subsequently embraced when articulated by a male colleague, thereby worsening the problem of unequal compensation (INVISIBILITY). About this, P24 commented: *"There have been instances where my ideas, which I proposed as part of a solution, were appropriated, even though when I initially spoke, nobody paid attention."* In addition, P37 added: *"men repeating what I say and getting the credit for it."*

Finally, in our analysis, we noted a potential linkage between the challenges encountered by women and the specific roles they assume in software development, as well as their involvement in project-related activities. Constructed from responses derived from the closed-ended questions, Figure 2 is founded on the challenges documented in the literature. As illustrated, women grapple with issues such as impostor syndrome, invisibility, stereotypes, and managing work-life balance regardless of their designated roles within the software team. However, a significant observation arises upon examining programmers and testers.

Female programmers face the most substantial struggles within toxic work environments and often face EXCLUSION FROM DECISIONS. In contrast, female software testers encounter LACK OF SUPPORT and experience their contributions being underestimated more frequently (UNDERVALUED WORK). The analysis of the open-ended questions supports these findings. We observed that those engaged in programming activities are prone to experiencing a sense of isolation and detachment from the team, concomitantly encountering pressure and frequent interruptions when attempting to communicate with their colleagues. On the other hand, half of the software testers in the sample reported that they are not taken seriously by the team when reporting failures or attempting to engage in discussions regarding implementation problems.

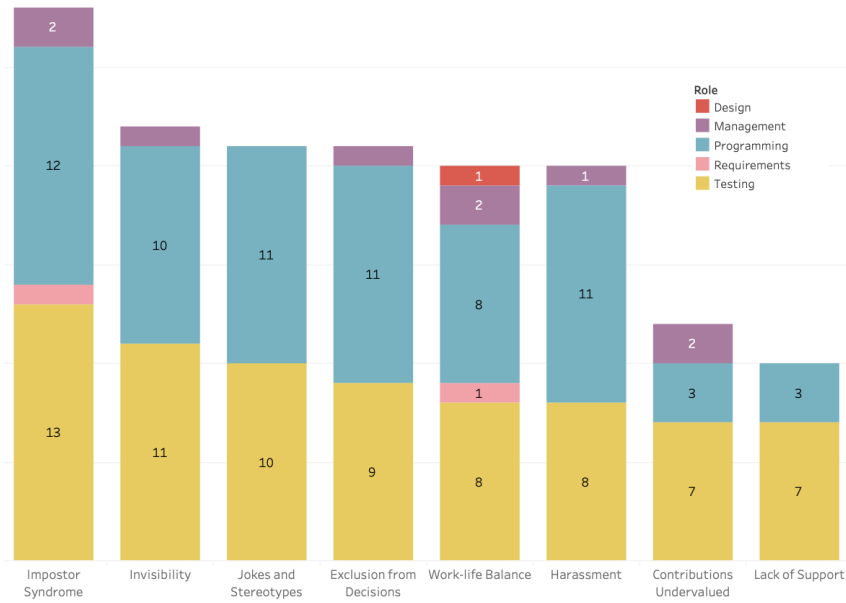

Figure 2: Challenges Distribution by Role in the Software Team

P23 is a programmer, and she shared *"always been very afraid of asking for help"* when she had doubts because she felt others would think she *"was incapable of performing the proposed tasks."* P05 is a software QA, and she reported that: *"male technical leaders don't easily accept bugs being pointed out in the developed code (…) my technical knowledge is frequently questioned."* P07 is a tester, and she added to this discussion: *"As a tester, I'm frequently not heard, and there have been times I suggested a potential bug root, only to be disregarded by the developer."*

*4.4.2 Strategies to Support Women in the Software Industry.*
As in Sections 4.2.2 and 4.3.2, we invited participants to share their thoughts on strategies to overcome the challenges they currently face in the software industry, aiming to identify ways to enhance their work experience. From their input, 11 specific actions emerged as having the potential to address the existing issues and foster a more gender-egalitarian software development environment. Table 6 summarizes these support strategies.

## 5 DISCUSSION

In this section, we present and compare our findings to existing literature, then delve into the path of women in software engineering. Finally, we acknowledge the study's limitations.

### 5.1 Enfolding the Literature

Our results indicate that the growing body of research dedicated to enhancing gender diversity in STEM fields holds promise for gradual improvement in this scenario. The challenges outlined for female software engineering students and professionals in our literature review were corroborated by our participants. Moreover, our study expanded upon the existing body of knowledge, highlighting novel obstacles, connecting these obstacles from different stages of female software engineers' careers (e.g., from the university to the industry and considering different roles in the software team), and proposing potential solutions that can be implemented in both academic and industrial settings. Our study also showed that many of the identified problems are cross-cultural, as some challenges identified in our study were highlighted in previous studies found in other contexts.

In Table 3, we summarized the challenges women face during their studies. The majority reported experiencing feeling insecure and fear of failure, thus facing difficulties while trying to fit in. Prior research suggests that, among engineering students, women tend to have worse mental health than men, in part because of additional stresses related to fitting in [4, 14, 30]. Aside from addressing systemic barriers and gendered disparities, universities should strive to promote mental health by tackling the normalization of high pressure and shame related to seeking help [4, 31]. This is achieved by fostering a culture of wellness through the actions of faculty and staff, such as introducing students to mental health content such as strategies for managing anxiety and encouraging professors to redesign curriculum to promote health and equity [4, 6, 29, 31, 46]. Normalizing failure by exposing students, particularly those belonging to underrepresented groups, to relatable career stories may also help reduce anxiety [8]. Tait et al. provided a comprehensive summary of proposed mental health interventions for engineering students [62].

Just under half of survey respondents felt they suffered from a lack of role models, jokes and stereotypes, and moral harassment. Research shows that lack of representation and stereotypes about both women and 'typical' programmers can discourage women from pursuing a career in software and hinder identification with the field [2, 12, 21, 52, 60, 64, 69]. Respondents recommended channels for reporting harassment, improving representation of women among faculty, showcasing women in the field, and exposing students to a wide range of career options. Two of the authors observe that they have incorporated the last two techniques in their own courses, to which students have responded positively (e.g., [16, 39]).

**Table 6: Solutions and Mitigation**

| Action | Description | Participant Quote |
|---|---|---|
| Diversity Awareness | Creating initiatives centered on raising the team's understanding of the vital role of diversity, particularly gender diversity, in advancing technologies. | "many initiatives towards enhancing gender equality often result in discussions and activities among women, with limited engagement from men, which is ineffective." - P19 |
| Female Leadership | Implementing equitable processes that enable females to attain leadership positions can motivate women within the company, fostering a sense of recognition and value | "establish initiatives to encourage more and more women to assume leadership positions." - P07 "(…) training females for leadership roles." - P21 |
| Team Empathy | Cultivate the team's ability to comprehend colleagues' limitations and acknowledge collective and individual contributions. | "Offer a supportive environment and never, under any circumstances, disregard or treat lightly any instance of harassment/abuse.." - P14 |
| Professional Development | Develop company programs designed to support the professional development of women, specifically targeting the enhancement of their technical skills. | "Allocate resources to training and professional development programs for women, as well as endorse community and educational initiatives." - P02 |
| Team Diversity | Improve team diversity as a tactic to introduce a diverse range of perspectives, including viewpoints on the significance of diversity and inclusion in the field of software engineering. | "Foster a diverse environment, starting in the recruitment process and extending it to the team composition." - P22 "Provide opportunities to people, no matter their background, e.g., gender, religion, age." - P16 |
| Team Inclusiveness | Develop practices to ensure women are valued, respected, and actively involved in the team. | "We need inclusion policies because many companies are hiring, but I have seen a lack of sensitive management of the challenges." - P15 |
| Unbiased Evaluation | Utilize fair evaluation methods that avoid bias against women while also recognizing the enduring obstacles they have confronted and their potential impact on outcomes. | "Perform comparisons of promotions and career paths between men and women in the same position, assessing if they follow similar trajectories and are evaluated equitably.." - P36 |
| Whistleblowing Structures | Establish a structure to facilitate reporting or disclosing wrongdoing, unethical behavior, or illegal activities within the team or the company. | "(…) anonymous channel for reporting harassment, enabling female employees to address uncomfortable situations." - P41 |
| Childcare Support | Provide assistance to mothers, assisting them in alleviating some of the pressures stemming from the challenges of balancing work and personal life. | "high-quality daycare support so that women can work without concerns." - P27 |
| Egalitarian Salary | Institute policies for salary equity. | "(…) compatible positions and salaries." - P08 |
| Misconduct Policies | Enforce strict penalties for confirmed instances of misconduct and harassment. | "(…) suppress sexist behaviors." - P28 |

As they join the workforce, distrust of their skills, stress caused by toxic environments, and biased hiring are the primary problems Brazilian women face (Table 4). From exclusion based on perceived incompetency to sexual harassment to biased job advertisements, all of these problems have been previously observed (e.g., [21, 65]). In transitioning to their first jobs, respondents found team empathy, other women in the team, technical help, and regular feedback helped improve the environment. Individual workers might be able to better their circumstances by unionizing to increase transparency and eliminate salary disparities [68]. For organizations to change, they can seek training from external organizations with relevant expertise, but it is essential that there is trust and willingness to collaborate among stakeholders [61]. Some methods which are associated with success are optional diversity training, cross-training and mentorship [18]. Organizations should also evaluate hiring practices which may lead to installing narcissistic and psychopathic leaders, such as hiring based on confidence rather than competence and preferencing charisma [11]. However, it is important to remember that any evaluations of competency which take place within a biased system may fail to accurately assess merit [9].

Table 5 gives the ongoing challenges faced by Brazilian women in the software industry. Impostor syndrome, invisibility, jokes and stereotypes, exclusion from decisions, work-life balance, and toxicity were particularly prevalent. Respondents provided a number of suggested actions for mitigating these problems, as shown in Table 6. As all of these challenges exist during the transition period as well, we have already positioned them in the literature. In the next section, we will discuss how various problems persist across different points of women's careers.

### 5.2 Navigating the Path of Women in Software Engineering

Figure 3 illustrates the trajectory women pursue when pursuing a career in software engineering. We see that some challenges, namely stereotypes, undervalued work, harassment, misconduct, and a missing sense of belonging, are present from the start of women's journeys and persist throughout their careers. Others, such as a lack of role models and segregation, affect the student years, while the shortage of opportunities appears to diminish as women become more experienced. Meanwhile, a host of new issues arise as women enter the workforce. Our participants also identified a number of strategies to address the challenges of each particular career stage.

One interesting facet of this research is that it shows that although some problems seem to develop later, they probably have antecedents in earlier periods. For example, while work-life balance emerges as a specific problem later in women's careers, we note that the culture of high pressure is already present during engineering studies, with students exposed to exceptionally high workloads, with observed problems such as FEAR OF FAILING. Several researchers have called for using the COVID-19 pandemic as an opportunity to re-imagine engineering studies and work to develop a more inclusive and mentally healthy environment [29, 50, 54].

Another thing worth noting is that the problems shown in Figure 3 are entirely systemic. These are not problems that women can easily address independently but rather require coordination and organizational commitment. This is true even for issues which, at first glance, appear to involve individuals (i.e., impostor syndrome). According to Mullangi and Jagsi, impostor syndrome is a symptom of structural inequalities. Like burnout, which is a response to an overwhelming and exhausting environment, impostor syndrome should be tackled by addressing the root causes. While impostor syndrome is widely experienced among professionals, minoritized individuals often suffer from it in greater numbers because observed inequalities may be attributed to personal shortcomings. Consequently, greater equality should lead to a lessening of impostor syndrome [41].

### 5.3 Threats to Validity

Our research targeted female software professionals exclusively and to ensure the relevance of our sample [37], we narrowed our focus to women involved in software development roles, encompassing positions like programmers, software testers, designers, and project managers, among others. We employed a convenience sampling strategy to directly reach out to these women, and we encouraged participants to extend the questionnaire solely to other women within the software industry, utilizing a snowball sampling approach.

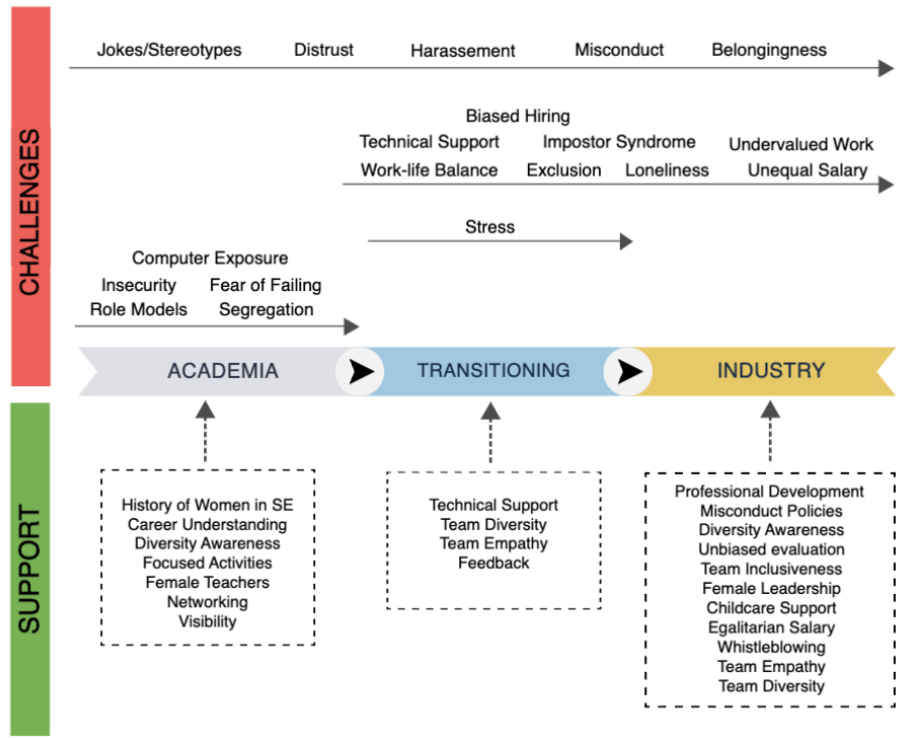

**Figure 3: Women's Journey in SE**

To ensure construct validity [37, 48] and eliminate potential confusion or ambiguities for participants, we conducted a validation of our pilot questionnaire with three individuals from the targeted population and who were not included in our final sample. Additionally, we strategically ordered our questions by presenting open-ended inquiries before closed-ended ones that drew upon literature-based evidence. This sequencing prevented participants from duplicating information already reported in prior studies and enabled us to introduce fresh perspectives (i.e., not captured in our literature review), thus enriching the body of knowledge surrounding this theme.

Finally, our survey is limited by its sample size and sampling strategies, which restricts statistical generalization to a broader population. However, a significant portion of our findings is rooted in qualitative exploration. Consequently, we anticipate that our insights can be extrapolated analytically to diverse contexts, i.e., educators and practitioners can learn from our narrative to plan strategies for addressing the gender gap in software engineering.

## 6 CONCLUSION

In this research, we conducted a cross-sectional survey to expand upon the current body of knowledge by investigating women's experiences in software engineering as they transitioned from the university to software companies. Employing this method, we gathered insights from a diverse group of 42 women engaged in various roles within the software development process, representing a broad spectrum of backgrounds and profiles in software engineering.

Our findings demonstrate that various challenges persist throughout women's careers in software engineering. From their university education to their work in the software industry, women struggle with gender biases, harassment, work-life imbalance, undervalued work, and low sense of belonging, in addition to individual challenges like impostor syndrome. Moreover, while certain challenges, such as a lack of role models and gender segregation, impact their student years, the scarcity of opportunities tends to decrease as their careers progress. Notably, we observed that challenges emerging later in their careers often have roots in earlier university experiences.

In summary, as we explored the path of women entering software engineering, we observed that many challenges are systemic and demand organizational commitment to address them effectively, even when considering individual issues, such as impostor syndrome, which is rooted in structural inequalities. Our study offers a set of recommendations to address these challenges and support the promotion of greater gender equality within the software engineering field.